\documentclass[journal]{IEEEtran}
\IEEEoverridecommandlockouts
\usepackage{cite}
\usepackage{amsmath,amssymb,amsfonts}
\usepackage[ruled,vlined]{algorithm2e}
\usepackage{graphicx}
\usepackage{textcomp}
\usepackage{xcolor}
\usepackage{booktabs}
\usepackage{multirow}
\usepackage{colortbl}
\usepackage{tabularx}
\definecolor{ourshl}{RGB}{232,222,246}

\renewcommand{\arraystretch}{1.08}

\newcommand{\resulttablesetup}{\footnotesize\setlength{\tabcolsep}{3.4pt}\renewcommand{\arraystretch}{1.10}}
\newcommand{\tb}[1]{\textbf{\boldmath #1}}
\def\BibTeX{{\rm B\kern-.05em{\sc i\kern-.025em b}\kern-.08em
    T\kern-.1667em\lower.7ex\hbox{E}\kern-.125emX}}
\begin{document}

\title{FORCE-Interior: Measurement-Consistent Adaptation of a Poisson-Flow Generative Prior for Interior CT}

\author{%
\IEEEauthorblockN{Kang Chen, Wenjun Xia, Member, IEEE,
Jianxu Wang, Mahmud Wasif Nafee, and Ge Wang, Life Fellow, IEEE}%
\thanks{Kang Chen is with the Department of Electrical, Computer,
and Systems Engineering, Rensselaer Polytechnic Institute,
Troy, NY 12180 USA (e-mail: chenk14@rpi.edu);
Wenjun Xia, Jianxu Wang, Mahmud Wasif Nafee, and Ge Wang are
with the Department of Biomedical Engineering, Rensselaer
Polytechnic Institute, Troy, NY 12180 USA. Wenjun Xia and Ge Wang are co-corresponding authors (e-mail: xwj90620@gmail.com; wangg6@rpi.edu).}%

}

\maketitle

\begin{abstract}
Interior tomography reconstructs a region of interest (ROI) from truncated projections, an ill-posed problem with non-unique solutions and truncation-induced bias. Existing deep-learning methods can be sensitive to changes in ROI geometry and noise, while expressive generative priors may produce measurement-inconsistent content without measurement constraints. We propose \textit{FORCE-Interior}, a training-free adaptation of a pretrained Poisson-flow generative prior to interior CT. A full-field-of-view (FOV) OS-SART warm start avoids forcing all measured attenuation into the ROI, and truncation-mask-aware OS-SART updates enforce data consistency throughout sampling. In our experiment, FORCE-Interior achieves the best PSNR, SSIM, and LPIPS at the two more severely truncated synthetic ROI sizes and competitive performance at the largest ROI, while maintaining low projection-domain residuals. These findings support the measurement-consistent adaptation of a reusable generative CT prior, while further clinical and patient-level validation remains necessary.

\end{abstract}


\section{Introduction}
Interior tomography aims to reconstruct a prescribed region of interest (ROI) from truncated projections~\cite{wang2013meaning}. 
This setting is particularly relevant when the target anatomy is spatially localized and ROI-focused reconstruction is desired.

However, the inverse problem of interior tomography has non-unique solutions and is fundamentally more challenging than standard CT because projection data are truncated, and the measured sinogram does not uniquely determine the attenuation map~\cite{yu2009compressed,wang2013meaning,kudo2013image}.
Unknown structures outside the ROI can still contribute to the measured rays, so different complete attenuation maps may be consistent with the same ROI-truncated data~\cite{paleo2017practical,wang2013meaning}.
Classical analytic reconstruction methods, such as standard filtered backprojection (FBP), which requires complete projection data, cannot be applied directly. Applying it to truncated measurements is mathematically inappropriate and can cause ambiguity or null-space artifacts in the reconstruction results~\cite{bilgot2011fbp,kudo2013image}. 

Representative conventional approaches include model-based algebraic iterative reconstruction methods such as the simultaneous algebraic reconstruction technique (SART)~\cite{andersen1984sart} and ordered-subset SART (OS-SART)~\cite{wang2004ossart}, but these methods can still exhibit artifacts and noise when projection data are truncated~\cite{kudo2013image,jin2012bone,zhang2009artifact,lu2013diffusion}. To further stabilize the reconstruction, regularized model-based methods incorporate manually designed priors or constraints into the inverse problem, including total variation minimization~\cite{sidky2008image}, compressed-sensing-based interior tomography~\cite{yu2009compressed}, and prior-image constrained compressed sensing (PICCS)~\cite{chen2008prior}.
Although these priors are interpretable, they are typically designed based on physical intuition or mathematical assumptions, which limits their ability to capture complex anatomical variability~\cite{arridge2019solving}. Moreover, denoising methods such as total variation (TV) may oversmooth anatomical structures and suppress fine details~\cite{Tian_2011}. 

Recently, many deep learning-based reconstruction methods have been studied to overcome the limitations of conventional reconstruction approaches and have shown promising performance. For example, CNN-based methods have improved artifact suppression for interior CT, particularly through image-domain, projection-domain, or both~\cite{han2018deeplearninginteriortomography,han2019networksolveroisdeep,han2025endtoenddeeplearninginterior}. For example, Han et al.~\cite{han2025endtoenddeeplearninginterior} developed an end-to-end dual-domain learning framework to jointly address projection-domain truncation and image-domain degradation in interior CT.
Compared with conventional iterative reconstruction methods, these approaches demonstrate stronger artifact suppression, improved visual quality, and faster inference. However, most such methods are deterministic mappings trained with paired data. Their performance can degrade under unseen ROI sizes, ROI positions, and different noise levels.

Diffusion-based and flow-based generative models~\cite{song2022solving,liu2022dolcemodelbasedprobabilisticdiffusion,guan2022generativemodelingsinogramdomain,xia2025tomographicfoundationmodel} provide flexible frameworks for CT inverse problems by learning expressive priors over high-quality CT images and recovering target images through iterative sampling. However, the flexibility of such expressive priors may also introduce reliability concerns. Without sufficient data-consistency constraints, generative sampling may produce anatomically plausible but measurement-inconsistent structures. A common strategy is to integrate the generative prior with physical data-consistency conditioning, balancing anatomical realism with measurement fidelity. For example, DOLCE~\cite{liu2022dolcemodelbasedprobabilisticdiffusion}, DDS~\cite{chung2024decomposeddiffusionsampleraccelerating}, and FORCE~\cite{xia2025tomographicfoundationmodel} demonstrate that data-consistency conditioning can guide sampling from expressive generative priors while preserving measurement fidelity.

Most generative CT reconstruction methods have been developed for low-dose, sparse-view, limited-angle, or metal-artifact settings~\cite{liu2022dolcemodelbasedprobabilisticdiffusion,guan2022generativemodelingsinogramdomain,xia2025tomographicfoundationmodel}. Recent work has also considered diffusion-bridge reconstruction from incomplete data, including truncated projections~\cite{wang2025projection}. Interior tomography nevertheless presents a distinct challenge: ROI-based truncation leaves low-frequency content and exterior attenuation non-identifiable. The question addressed here is whether a reusable pretrained Poisson-flow CT prior can be adapted without retraining specific to interior tomography while remaining explicitly consistent with the measured truncated sinogram.

To address these problems, we propose \textit{\textbf{FORCE-Interior}}, which adapts a pretrained Poisson-flow generative prior to interior tomography without retraining specific to interior tomography. FORCE-Interior first constructs a full-FOV warm start that avoids forcing all measured attenuation into the ROI and then enforces per-step OS-SART consistency with the measured truncated sinogram. Experiments demonstrate improved structural and perceptual reconstruction quality over representative conventional, CNN-based, and diffusion-based baselines in most ROI-based evaluations.

The main contributions are summarized as follows:

\begin{itemize}

    \item We formulate interior tomography as a training-free adaptation problem, using a pretrained Poisson-flow (PFGM++) CT prior without retraining it on interior measurements.

    \item We introduce an interior-specific conditioning framework that combines full-FOV OS-SART initialization with truncation-mask-aware, per-step data consistency, reducing the bias caused by forcing exterior attenuation into an ROI-restricted image.

    \item We evaluate structural, perceptual, projection-domain, and lesion-retention behavior on controlled synthetic data and real scanner measurements across ROI sizes, positions, and dose levels.

\end{itemize}

\section{Preliminaries}
\label{sec:preliminary}
\subsection{Elucidating the Design Space of Diffusion-Based Generative Models}
Let $\mathbf{x}_0  \sim p_{\mathrm{data}}(\mathbf{x})$ denote a clean image sampled from the data distribution. In diffusion-based generative models, the forward noising process can be formulated as a stochastic differential equation (SDE):
\begin{equation}
\mathrm{d}\mathbf{x}=f(\mathbf{x}, t)\,\mathrm{d}t+g(t)\,\mathrm{d}\mathbf{w},
\label{eq:forward_sde}
\end{equation}
where $\mathbf{x}$ denotes the image state at time $t$, $f(\mathbf{x}, t)$ is the drift term describing the deterministic evolution of the state, $g(t)$ is the diffusion coefficient controlling the noise scale, and $\mathrm{d}\mathbf{w}$ denotes the increment of a standard Wiener process. This forward SDE progressively corrupts $\mathbf{x}_0$ by injecting noise, transforming the data distribution into an isotropic Gaussian distribution. The reverse process removes the injected noise and recovers samples from the original data distribution. Different formulations vary in how they solve this reverse process. For example, DDPMs~\cite{ho2020denoisingdiffusionprobabilisticmodels} learn a discrete reverse denoising transition $p_{\theta}(\mathbf{x}_{t-1}\mid \mathbf{x}_{t})$ by matching the forward-process posterior $q(\mathbf{x}_{t-1}\mid \mathbf{x}_{t}, \mathbf{x}_{0})$, while score-based diffusion models~\cite{song2021score} learn the time-dependent score $s_{\theta}(\mathbf{x}_{t}, t) \approx \nabla_{\mathbf{x}_{t}} \log p_{t}(\mathbf{x}_{t})$.

EDM further unifies these perspectives through a denoising-based score parameterization, noise-dependent network preconditioning, weighted training objectives, and flexible ODE samplers, providing a systematic design space for diffusion-based generative modeling~\cite{karras2022elucidating}. In EDM, the probability flow ODE can be written as
\begin{equation}
\mathrm{d}\mathbf{x}
=
-\dot{\sigma}(t)\sigma(t)
\nabla_{\mathbf{x}}\log p(\mathbf{x};\sigma(t))\,\mathrm{d}t,
\label{eq:edm_probability_flow}
\end{equation}
where $\sigma(t)$ denotes the noise schedule, $\dot{\sigma}(t)$ is its time derivative, and $\nabla_{\mathbf{x}}\log p(\mathbf{x};\sigma(t))$ is the score function of the noise-perturbed data distribution. This ODE describes the deterministic reverse trajectory that moves a noisy sample toward the data distribution. Instead of directly training a network to predict the score, EDM trains a denoising network $f_{\theta}$ to recover a clean image from its noisy observation:
\begin{equation}
\resizebox{.95\columnwidth}{!}{$
\mathcal{L}_{\mathrm{EDM}}(\theta)
=
\mathbb{E}_{\mathbf{x}_0\sim p(\mathbf{x}_0)}
\mathbb{E}_{\sigma\sim p(\sigma)}
\mathbb{E}_{\boldsymbol{\epsilon}\sim\mathcal{N}(\mathbf{0},\mathbf{I})}
\lambda_{\mathrm{EDM}}(\sigma)
\left\|
f_{\theta}(\mathbf{x}_0+\sigma\boldsymbol{\epsilon};\sigma)
-
\mathbf{x}_0
\right\|_2^2
$}
\label{eq:edm_loss}
\end{equation}
Here, $\mathbf{x}_0$ is a clean training sample, $\boldsymbol{\epsilon}$ is Gaussian noise, $\sigma$ controls the noise level, and $\lambda_{\mathrm{EDM}}(\sigma)$ is the noise-dependent EDM loss weight. The learned denoiser can then be converted into the score function by
\begin{equation}
\nabla_{\mathbf{x}}\log p(\mathbf{x};\sigma)
=
\frac{f_{\theta}(\mathbf{x};\sigma)-\mathbf{x}}{\sigma^2}.
\label{eq:edm_score}
\end{equation}
Therefore, EDM converts score estimation into a denoising problem, providing a convenient foundation for combining diffusion-based priors with other generative-flow formulations.

\subsection{EDM-based Poisson Flow Generative Models}
Generative modeling can also be viewed as a thermodynamic problem, whereas the Poisson flow generative model (PFGM)~\cite{xu2022poisson} reformulates generative modeling as an electrostatic field problem. The PFGM forward process transforms the original charge distribution into a uniform charge distribution on a hemisphere by following the ordinary differential equation
\begin{equation}
    \mathrm{d}\tilde{\mathbf{x}} = \mathbf{E}(\tilde{\mathbf{x}})\,\mathrm{d}t,
    \label{eq:pfgm_ode}
\end{equation}
where $\tilde{\mathbf{x}}=(\mathbf{x},\mathbf{z})\in\mathbb{R}^{N+D}$ denotes the augmented state, $\mathbf{z}\in\mathbb{R}^{D}$ is the augmented variable, and $\mathbf{E}(\tilde{\mathbf{x}})$ is the electric field induced by the data distribution. Specifically, the field is defined as
{\footnotesize
\begin{equation}
\hspace*{-0.8em}
\mathbf{E}(\tilde{\mathbf{x}})
=
\frac{1}{S^{N+D-1}(1)}
\int
\frac{\tilde{\mathbf{x}}-\tilde{\mathbf{x}}_0}
{\left\|\tilde{\mathbf{x}}-\tilde{\mathbf{x}}_0\right\|^{N+D}}
p(\mathbf{x}_0)\,\mathrm{d}\mathbf{x}_0,
\label{eq:pfgm_field}
\end{equation}
}
where $\tilde{\mathbf{x}}_0=(\mathbf{x}_0,\mathbf{0})$ is the augmented representation of a data sample $\mathbf{x}_0$, $p(\mathbf{x}_0)$ is the data density, and $S^{N+D-1}(1)$ denotes the surface area of the unit $(N+D-1)$-sphere. Under this formulation, samples are generated by solving the reverse trajectory from the hemisphere prior back to the data manifold.

To combine EDM with PFGM++~\cite{xu2023pfgmpp}, the EDM noise scale $\sigma$ is mapped to the radial coordinate of the augmented PFGM++ space. Let $r(\tilde{\mathbf{x}})=\|\mathbf{z}\|_2$ denote the radius of an augmented data point. Since $\|\boldsymbol{\epsilon}\|_2 \approx \sqrt{D}$ for $\boldsymbol{\epsilon}\sim\mathcal{N}(\mathbf{0},\mathbf{I}_D)$, the PFGM++ radius is set as $r=\sigma\sqrt{D}$.
This mapping aligns the EDM noise level with the radial coordinate in the PFGM++ hyperspherical space. Thus, the Gaussian perturbation in EDM can be replaced by a Poisson-flow perturbation on the hypersphere. Following this mapping, the EDM-based PFGM++ training objective can be written as
\begin{equation}
\resizebox{.95\columnwidth}{!}{$
\mathcal{L}(\theta)
=
\mathbb{E}_{\mathbf{x}_0\sim p(\mathbf{x}_0)}
\mathbb{E}_{\sigma\sim p(\sigma)}
\mathbb{E}_{R\sim p(R\mid\sigma),\,\mathbf{v}\sim U(S^{N-1})}
\lambda_{\mathrm{EDM}}(\sigma)
\left\|
f_{\theta}(\mathbf{x}_0+R\mathbf{v};\sigma)
-
\mathbf{x}_0
\right\|_2^2
$}
\label{eq:edm_pfgm_loss}
\end{equation}
where $\mathbf{v}$ is a unit direction sampled uniformly from the image-space sphere $S^{N-1}$ and the radius follows the PFGM++ radial law $R=r\sqrt{\beta/(1-\beta)}$ with $\beta\sim\mathrm{Beta}(N/2,D/2)$, i.e., the image-space displacement obtained by sampling uniformly on the shell of radius $r$ in the augmented space. Compared with the EDM objective in Eq.~\eqref{eq:edm_loss}, Eq.~\eqref{eq:edm_pfgm_loss} preserves the same denoising target but replaces the Gaussian perturbation $\sigma\boldsymbol{\epsilon}$ with the hyperspherical perturbation $R\mathbf{v}$, which converges to the Gaussian one as $D\to\infty$. The overall training procedure is summarized in Algorithm~\ref{alg:pfgm_train}.

\begin{algorithm}[t]
\caption{EDM-based PFGM++ training}
\label{alg:pfgm_train}
\KwIn{Dataset $p(\mathbf{x}_0)$, noise schedule $p(\sigma)$}
\KwOut{Trained model $\mathbf{f}_{\boldsymbol{\theta}}$}
\While{not converged}{
    $\mathbf{x}_0\sim p(\mathbf{x}_0)$, $\sigma\sim p(\sigma)$\;
    
    $r=\sigma\sqrt{D},\quad
    \beta\sim\mathrm{Beta}(N/2,D/2)$\;
    
    $R=r\sqrt{\beta/(1-\beta)}$\;
    
    $\mathbf{u}\sim\mathcal{N}(\mathbf{0},\boldsymbol{I})$, $\mathbf{v}=\mathbf{u}/\|\mathbf{u}\|_2$\;
    $\mathcal{L}(\boldsymbol{\theta})=\lambda_{\mathrm{EDM}}(\sigma)\|\mathbf{f}_{\boldsymbol{\theta}}(\mathbf{x}_0+R\mathbf{v};\sigma)-\mathbf{x}_0\|_2^2$\;
    Update $\boldsymbol{\theta}$\;
}
\end{algorithm}

\subsection{FORCE: Foundation Model for CT Reconstruction}

Computed tomography (CT) reconstruction is a classical inverse problem that aims to recover an image $\mathbf{x}$ from projection measurements $\mathbf{b}$. The measurement model can be written as
\begin{equation}
    \mathbf{b}=\mathbf{H}\mathbf{x}+\mathbf{n},
    \label{eq:ct_forward_model}
\end{equation}
where $\mathbf{H}$ denotes the system's forward projection matrix, and $\mathbf{n}$ represents measurement noise. From a Bayesian perspective, model-based iterative reconstruction (MBIR) estimates $\mathbf{x}$ by maximizing the posterior distribution $p(\mathbf{x}\mid \mathbf{b})$. This leads to the following variational formulation:
\begin{equation}
    \hat{\mathbf{x}}
    =
    \arg\min_{\mathbf{x}}
    \frac{1}{2}
    \left\|
    \mathbf{H}\mathbf{x}-\mathbf{b}
    \right\|_2^2
    +
    \lambda \mathcal{R}(\mathbf{x}),
    \label{eq:mbir_objective}
\end{equation}
where the first term enforces data fidelity to the measured projection data, $\mathcal{R}(\mathbf{x})$ is an image prior that regularizes the reconstruction, and $\lambda$ controls the trade-off between the two terms.
FORCE~\cite{xia2025tomographicfoundationmodel} first learns the distribution of normal-dose CT images through an EDM-based PFGM++. During inference, this learned prior guides the sampling trajectory toward the manifold of realistic CT images, while task-specific data-fidelity conditioning ensures consistency with the measured projection data. In practice, different CT reconstruction tasks require different conditioning strategies. FORCE also incorporates total variation (TV) regularization to suppress structural artifacts and uses momentum to accelerate sampling. The overall sampling procedure is summarized in Algorithm~\ref{alg:force_sampling}.

\begin{algorithm}[t]
\caption{FORCE Sampling}
\label{alg:force_sampling}
\KwIn{Corrupted projection data $\mathbf{p}$, time schedule $\{t_i\}_{i=0}^{T}$}
\KwOut{Reconstructed image $\mathbf{x}_T$}
Initialize with a low-quality image $\mathbf{x}_{\mathrm{init}}$ from $\mathbf{p}$\;
$\mathbf{u}\sim\mathcal{N}(\mathbf{0},\boldsymbol{I})$, $\mathbf{v}=\mathbf{u}/\|\mathbf{u}\|_2$\;
$\beta\sim\mathrm{Beta}(N/2,D/2)$, $R_0=\sigma(t_0)\sqrt{D}\sqrt{\beta/(1-\beta)}$\;
Initialize sampling: $\mathbf{x}_0=\mathbf{x}_{\mathrm{init}}+R_0\cdot\mathbf{v}$\;
$\xi_{-1}=1$, $\hat{\mathbf{x}}_{-1}=\mathbf{x}_{\mathrm{init}}$\;
\For{$i=0$ \KwTo $T-1$}{
    $\xi_i=\dfrac{1+\sqrt{1+4\xi_{i-1}^2}}{2}$\;
    $\bar{\mathbf{x}}_i=\mathrm{Conditioning}(\mathbf{x}_i,\mathbf{p})$\;
    $\hat{\mathbf{x}}_i=\arg\min_{\mathbf{z}}\frac{1}{2}\|\mathbf{z}-\mathbf{f}_{\boldsymbol{\theta}}(\bar{\mathbf{x}}_i;\sigma_i)\|_2^2+\lambda\cdot\mathrm{TV}(\mathbf{z})$\;
    $\hat{\mathbf{x}}_i\leftarrow\hat{\mathbf{x}}_i+\dfrac{\xi_{i-1}-1}{\xi_i}(\hat{\mathbf{x}}_i-\hat{\mathbf{x}}_{i-1})$\;
    $\mathbf{d}_i=(\mathbf{x}_i-\hat{\mathbf{x}}_i)/\sigma_i$\;
    $\mathbf{x}_{i+1}=\mathbf{x}_i+(t_{i+1}-t_i)\mathbf{d}_i$\;
  }

\end{algorithm}

\section{Methodology}
\subsection{Problem Formulation of Interior Tomography}
Let $\boldsymbol{\mu}$ denote the attenuation image and $I_{0,j}$ the incident photon count for ray $j$. The photon-counting model used in our simulations is
\begin{equation}
\begin{aligned}
\mathbf{q} &\sim \mathrm{Poisson}\!\left(\mathbf{I}_0 \odot \exp(-\mathcal{A}\boldsymbol{\mu})\right),\\
\mathbf{y} &= \mathbf{M}\left[-\log\!\left(\frac{\max(\mathbf{q},1)}{\mathbf{I}_0}\right)\right],
\end{aligned}
\label{eq:photon_model}
\end{equation}
where $\mathcal{A}\in\mathbb{R}^{P\times N}$ maps an $N$-pixel image to $P=V\times B$ line integrals, $\odot$ denotes elementwise multiplication, and $\mathbf{M}=\operatorname{diag}(\mathbf{m})$ is the view-dependent binary truncation mask. For an eccentric ROI, its support is forward projected at each view and $\mathbf{M}$ retains the detector interval intersecting that projected support. Hence the retained interval can shift with view angle. Thus, the reconstruction can be generally written as
\begin{equation}
\hat{\boldsymbol{\mu}} = \arg\min_{\boldsymbol{\mu}} \frac{1}{2} \left\| \mathbf{M}\mathcal{A}\boldsymbol{\mu} - \mathbf{y} \right\|_{\mathbf{W}}^2 + \lambda \mathcal{R}(\boldsymbol{\mu}), \label{eq:interior_inverse}
\end{equation}
where $\mathbf{W}$ may encode ray-dependent statistical weights and $\mathcal{R}$ denotes a generic image prior. Our current OS-SART implementation uses uniform weights, an approximation to the Poisson likelihood and a limitation at very low photon counts.

\subsection{Solving Interior Tomography with FORCE}
We solve interior tomography using the pretrained PFGM++ model of Section~\ref{sec:preliminary} as an implicit prior, without retraining specific to interior tomography. Sampling integrates the EDM probability-flow ODE in Eq.~\eqref{eq:edm_probability_flow} backward, starting from a measurement-driven warm start perturbed to an intermediate noise level $t_{\mathrm{start}}$. At every step $i$ with noise level $\sigma_i=\sigma(t_i)$, the denoiser yields
\begin{equation}
\tilde{\mathbf{x}}_{0,i} = f_{\theta}(\mathbf{x}_i;\sigma_i),
\label{eq:force_denoise}
\end{equation}
into which we inject data consistency before taking the ODE step. The full procedure is summarized in Algorithm~\ref{alg:force_interior}.

\textbf{Full field-of-view warm start.}
Following FORCE, sampling starts from a measurement-driven OS-SART reconstruction, $\mathbf{x}_{\mathrm{init}}$. A ray intersecting the ROI can also traverse tissue outside it~\cite{paleo2017practical}. In the ROI-restricted control, pixels outside the ROI are fixed to zero and only interior pixels are updated. Exterior attenuation may then be incorrectly attributed to ROI pixels, producing intensity bias and cupping. Updating the full grid instead distributes measured attenuation over a physically plausible support and yields a less biased ROI initialization. It does not make the exterior uniquely identifiable from truncated data. The initialization is perturbed to an intermediate noise level. Here $t_{\mathrm{start}}\in(0,1]$ indexes the selected starting point on the full EDM schedule. At $t_{\mathrm{start}}{=}1$, the perturbation is maximal and the warm start has little influence, but the state is not literally pure noise because $\mathbf{x}_{\mathrm{init}}$ remains present. Let $t_0$ be the first node of the resulting reverse schedule,
\begin{equation}
\mathbf{x}_{t_0}=\mathbf{x}_{\mathrm{init}}+R_0\,\mathbf{v},\qquad \mathbf{v}\sim U(S^{N-1}),
\label{eq:warmstart}
\end{equation}
where $R_0=\sigma(t_0)\sqrt{D}\,\sqrt{\beta/(1-\beta)}$ with $\beta\sim\mathrm{Beta}(N/2,D/2)$ is the same PFGM++ perturbation used during training (Algorithm~\ref{alg:pfgm_train}), applied at noise level $\sigma(t_0)$. This keeps the measurement-constrained structure of $\mathbf{x}_{\mathrm{init}}$ while giving the prior room to fix the residual low-frequency bias and sharpen detail.

\textbf{Per-step data-consistency conditioning.}
At each reverse step, after the denoiser produces $\tilde{\mathbf{x}}_{0,i}$ in Eq.~\eqref{eq:force_denoise}, we obtain a distinct conditioned estimate $\bar{\mathbf{x}}_{0,i}$ by re-imposing agreement with the ROI-truncated sinogram $\mathbf{y}$ using OS-SART. Specifically, the $V$ projection views are partitioned into $G$ ordered subsets $\{\mathcal{S}_g\}_{g=1}^{G}$, and the estimate is refined subset by subset over the full image grid:
\begin{equation}
\label{eq:ossart}
\bar{\mathbf{x}}_{0,i}
\leftarrow
\tilde{\mathbf{x}}_{0,i}
+
\omega \mathbf{C}_g
(\mathbf{M}_g\mathcal{A}_{\mathcal{S}_g})^{\top}
\mathbf{R}_g
\left(
\mathbf{y}_{\mathcal{S}_g}
-
\mathbf{M}_g\mathcal{A}_{\mathcal{S}_g}\tilde{\mathbf{x}}_{0,i}
\right),
\end{equation}
where $g=1,\ldots,G$, $\mathcal{A}_{\mathcal{S}_g}\in\mathbb{R}^{P_g\times N}$ stacks the projection rows of the views in $\mathcal{S}_g$ (with $P_g$ the number of measurements in the subset), $\mathbf{M}_g\in\{0,1\}^{P_g\times P_g}$ denotes the diagonal truncation mask associated with the $g$-th subset, $\mathbf{y}_{\mathcal{S}_g}\in\mathbb{R}^{P_g}$ is the corresponding block of $\mathbf{y}$, and
$\mathbf{R}_g=\operatorname{diag}\left(1/(\mathbf{M}_g\mathcal{A}_{\mathcal{S}_g}\mathbf{1}+\varepsilon)\right)\in\mathbb{R}^{P_g\times P_g}$
and $\mathbf{C}_g=\operatorname{diag}\left(1/((\mathbf{M}_g\mathcal{A}_{\mathcal{S}_g})^{\top}\mathbf{1}+\varepsilon)\right)\in\mathbb{R}^{N\times N}$ are the projection- and image-domain ray-length normalization matrices, respectively, with $\mathbf{1}$ the all-ones vector of matching dimension. Here, $\omega$ is a relaxation factor and $\varepsilon$ is a small constant used to avoid division by zero. This update anchors the reconstruction to the measured ROI-truncated projections, while the generative prior regularizes the global and low-frequency components that are ambiguous under ROI truncation. Since direct OS-SART correction does not explicitly model measurement noise, repeated updates may amplify noise in noisy acquisitions. To mitigate this 
effect, we apply a lightweight total-variation (TV) proximal step $\mathrm{prox}_{\lambda\,\mathrm{TV}}$~\cite{chambolle2011first} following the native FORCE sampler. The conditioned estimate is then propagated by a standard Heun ODE step. In $\mathrm{HeunStep}$, the corrector re-evaluates $\mathbf{f}_{\boldsymbol{\theta}}$ at the predicted state and $t_{i+1}$. OS-SART and TV are applied to the predictor evaluation only and are not repeated in the corrector. The complete procedure is summarized in Algorithm~\ref{alg:force_interior}.

\begin{algorithm}[t]
\caption{FORCE-Interior reconstruction}
\label{alg:force_interior}
\KwIn{ROI-truncated sinogram $\mathbf{y}$, operator $\mathbf{M}\mathcal{A}$, prior $\mathbf{f}_{\boldsymbol{\theta}}$, time schedule $\{t_i\}_{i=0}^{T}$, subsets $G$, TV weight $\lambda$}
\KwOut{Reconstructed image $\mathbf{x}$}
$\mathbf{x}_{\mathrm{init}}\leftarrow \mathrm{OS\text{-}SART}_{\mathrm{full\text{-}FOV}}(\mathbf{y})$\tcp*{warm start}
$\mathbf{v}\sim U(S^{N-1})$, $\beta\sim\mathrm{Beta}(N/2,D/2)$, $R_0=\sigma(t_0)\sqrt{D}\sqrt{\beta/(1-\beta)}$\;
$\mathbf{x}\leftarrow \mathbf{x}_{\mathrm{init}}+R_0\,\mathbf{v}$\tcp*{Eq.~\eqref{eq:warmstart}}
\For{$i=0$ \KwTo $T-1$}{
  $\tilde{\mathbf{x}}_{0,i} \leftarrow \mathbf{f}_{\boldsymbol{\theta}}(\mathbf{x};\sigma(t_i))$\tcp*{denoise, Eq.~\eqref{eq:force_denoise}}
  $\bar{\mathbf{x}}_{0,i} \leftarrow \mathrm{OS\text{-}SART}(\tilde{\mathbf{x}}_{0,i},\mathbf{y};G)$\tcp*{Eq.~\eqref{eq:ossart}}
  $\bar{\mathbf{x}}_{0,i} \leftarrow \mathrm{prox}_{\lambda\,\mathrm{TV}}(\bar{\mathbf{x}}_{0,i})$\tcp*{TV regularization}
  $\mathbf{x}\leftarrow \mathrm{HeunStep}(\mathbf{x},\bar{\mathbf{x}}_{0,i},t_i,t_{i+1})$\tcp*{ODE step}
}
\Return $\mathbf{x}$
\end{algorithm}

\section{Experiments}
\label{sec:experiments}

\subsection{Datasets and Evaluation Protocol}
\label{sec:protocol}

We conducted all experiments on two public datasets. The first is the Low Dose CT Image and Projection Data (LDCT-and-Projection-data) collection from TCIA~\cite{moen2021ldct}, which provides raw scanner projection data with clinical annotations. The second is the 2016 NIH-AAPM-Mayo Clinic Low-Dose CT Grand Challenge~\cite{mccollough2017low}, which provides full-dose reconstructed images that serve as the ground truth for the synthetic evaluation. We used 1,923 slices from eight patients for the PFGM++ prior training and the remaining 445 slices from two held-out patients of the second dataset for synthetic evaluations.
Following the standard interior formulation, we generated truncated measurements using view-dependent detector masks obtained by forward-projecting prescribed centered or eccentric disk ROIs. ROI centers were fixed before evaluation. Only synthetic-lesion locations were randomly drawn.
For synthetic evaluations, we further injected Poisson noise into the truncated measurements using a physical photon-counting model at a dose of $I_0{=}10^{6}$ photons per detector bin for all baselines. In this setting, the held-out full-dose image serves as the ground truth, following the protocol of prior generative CT reconstruction work, and the truncated measurements are simulated from it, so all metrics are computed against a true reference.
For real scanner evaluations, we ran the experiment on both full-dose and low-dose data. Because real acquisitions provide no independent ground truth, we follow the standard low-dose CT protocol and use the vendor full-dose reconstruction as the reference rather than as ground truth. The rebinning ceiling obtained from the untruncated sinogram is described in Supplementary Sec.~I.
We report results on a prespecified structure-rich ROI subset, with its selection criterion and radius-dependent sample sizes described in Supplementary Sec.~I. Because eligible slices differ across radii, cross-radius trends are interpreted descriptively rather than as paired causal effects of ROI size.
All experiments were performed on a single NVIDIA RTX A4000 GPU. FORCE-Interior reconstructed one slice in approximately 42~s, with a peak VRAM usage of about 3.5~GB. Full geometry, solver, and PFGM++ prior training settings are listed in Supplementary Sec.~I.

\begin{table*}[!t]
\centering
\caption{Reconstruction on real scanner projection data at $r\in\{96,128,160\}$~px (ROI, $[0,2000]$~HU, 50 slices from five patients). Best in \textbf{bold}, second best \underline{underlined}. $^{*}$/$^{**}$ denote  slice-level Holm-adjusted paired Wilcoxon $p<0.05$/$p<0.001$. Correlated slices are not independent, and patient-level results are reported in Supplementary Sec.~V.}
\label{tab:real_results}
\scriptsize
\setlength{\tabcolsep}{2pt}
\renewcommand{\arraystretch}{1.12}
\begin{tabularx}{\textwidth}{c l *{6}{>{\centering\arraybackslash}X}}
\toprule
\multicolumn{2}{c}{\textit{Experimental setting}}
& \multicolumn{3}{c}{\textit{Normal dose}}
& \multicolumn{3}{c}{\textit{Low dose}} \\
\cmidrule(lr){1-2} \cmidrule(lr){3-5} \cmidrule(lr){6-8}
ROI radius
& Method
& PSNR $\uparrow$
& SSIM $\uparrow$
& LPIPS $\downarrow$
& PSNR $\uparrow$
& SSIM $\uparrow$
& LPIPS $\downarrow$ \\
\midrule

\multirow{4}{*}{$96$} & OS-SART
& $22.84 \pm 3.43$ & $0.951 \pm 0.014$ & $0.278 \pm 0.047$
& $22.83 \pm 3.47$ & $0.945 \pm 0.015$ & $0.342 \pm 0.056$ \\

 & SART-TV
& $22.93 \pm 3.44$ & \underline{$0.953 \pm 0.013$} & \underline{$0.239 \pm 0.045$}
& $23.06 \pm 3.56$ & \underline{$0.950 \pm 0.015$} & \underline{$0.287 \pm 0.057$} \\

 & DDS
& \underline{$28.24 \pm 3.02$} & $0.941 \pm 0.022$ & $0.301 \pm 0.086$
& \underline{$27.82 \pm 2.48$} & $0.941 \pm 0.019$ & $0.343 \pm 0.079$ \\

\rowcolor{ourshl}\cellcolor{white}
 & \textbf{Ours}
& \tb{$28.89 \pm 2.46$} & \tb{$0.959 \pm 0.015$} & \tb{$0.189 \pm 0.057$}$^{**}$
& \tb{$28.86 \pm 2.28$}$^{*}$ & \tb{$0.958 \pm 0.013$}$^{*}$ & \tb{$0.239 \pm 0.057$}$^{*}$ \\

\cmidrule(lr){1-8}

\multirow{5}{*}{$128$} & OS-SART
& $28.74 \pm 1.69$ & $0.931 \pm 0.018$ & $0.247 \pm 0.035$
& $28.48 \pm 1.68$ & $0.923 \pm 0.012$ & $0.299 \pm 0.039$ \\

 & SART-TV
& $28.79 \pm 1.61$ & \underline{$0.931 \pm 0.021$} & \underline{$0.212 \pm 0.043$}
& $28.71 \pm 1.60$ & \underline{$0.926 \pm 0.017$} & \underline{$0.256 \pm 0.042$} \\

 & ROI-CT-CNN
& $28.12 \pm 1.73$ & $0.913 \pm 0.021$ & $0.358 \pm 0.037$
& $27.91 \pm 1.48$ & $0.908 \pm 0.016$ & $0.346 \pm 0.040$ \\

 & DDS
& \underline{$30.69 \pm 2.44$} & $0.930 \pm 0.029$ & $0.273 \pm 0.075$
& \underline{$30.00 \pm 1.99$} & $0.924 \pm 0.024$ & $0.289 \pm 0.075$ \\

\rowcolor{ourshl}\cellcolor{white}
 & \textbf{Ours}
& \tb{$31.84 \pm 1.84$}$^{**}$ & \tb{$0.947 \pm 0.024$}$^{**}$ & \tb{$0.172 \pm 0.049$}$^{**}$
& \tb{$31.16 \pm 1.50$}$^{**}$ & \tb{$0.939 \pm 0.019$}$^{**}$ & \tb{$0.225 \pm 0.046$}$^{**}$ \\

\cmidrule(lr){1-8}

\multirow{4}{*}{$160$} & OS-SART
& $34.60 \pm 2.41$ & $0.945 \pm 0.026$ & $0.181 \pm 0.025$
& $32.90 \pm 1.80$ & $0.927 \pm 0.022$ & $0.230 \pm 0.033$ \\

 & SART-TV
& \underline{$34.90 \pm 2.60$} & \underline{$0.947 \pm 0.029$} & \underline{$0.140 \pm 0.025$}
& \underline{$33.74 \pm 2.16$} & \underline{$0.935 \pm 0.026$} & \tb{$0.176 \pm 0.029$} \\

 & DDS
& $34.46 \pm 1.68$ & $0.945 \pm 0.020$ & $0.225 \pm 0.047$
& $33.22 \pm 1.45$ & $0.930 \pm 0.018$ & $0.205 \pm 0.028$ \\

\rowcolor{ourshl}\cellcolor{white}
 & \textbf{Ours}
& \tb{$35.58 \pm 2.66$}$^{**}$ & \tb{$0.956 \pm 0.028$}$^{**}$ & \tb{$0.127 \pm 0.027$}$^{**}$
& \tb{$34.07 \pm 2.08$}$^{*}$ & \tb{$0.940 \pm 0.024$}$^{**}$ & \tb{$0.176 \pm 0.029$} \\

\bottomrule
\end{tabularx}
\end{table*}

\begin{figure*}[!t]
\centering
\includegraphics[width=0.92\textwidth]{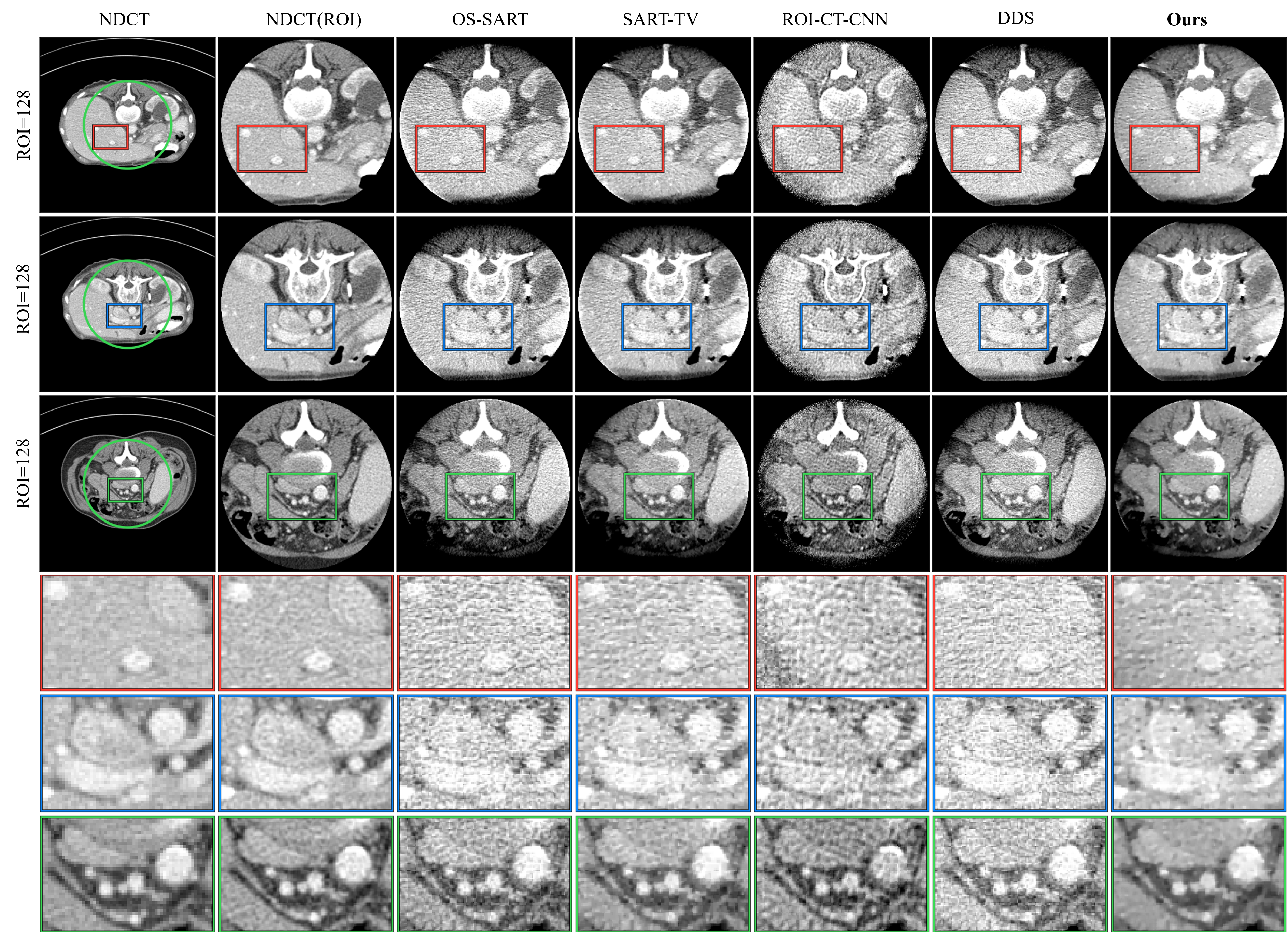}
\caption{Qualitative comparison on real scanner data (low dose, centred ROI, $r{=}128$~px), soft-tissue window $[-160,240]$~HU.}
\label{fig:real_centered}
\end{figure*}

\subsubsection{Metrics}
We evaluate reconstruction quality using three metrics. The structural similarity index measure (SSIM)~\cite{wang2004image} assesses whether anatomical structures, edges, and textures in the reconstructed slices are consistent with the reference. The peak signal-to-noise ratio (PSNR) measures pixel-level error in Hounsfield unit (HU) values. The learned perceptual image patch similarity (LPIPS)~\cite{zhang2018unreasonable} is a deep-feature-based perceptual distance. Higher PSNR and SSIM and lower LPIPS indicate better agreement with the reference. All quantitative metrics are computed within the ROI in the $[0,2000]$~HU window.

\begin{figure*}[!t]
\centering
\includegraphics[width=0.80\textwidth]{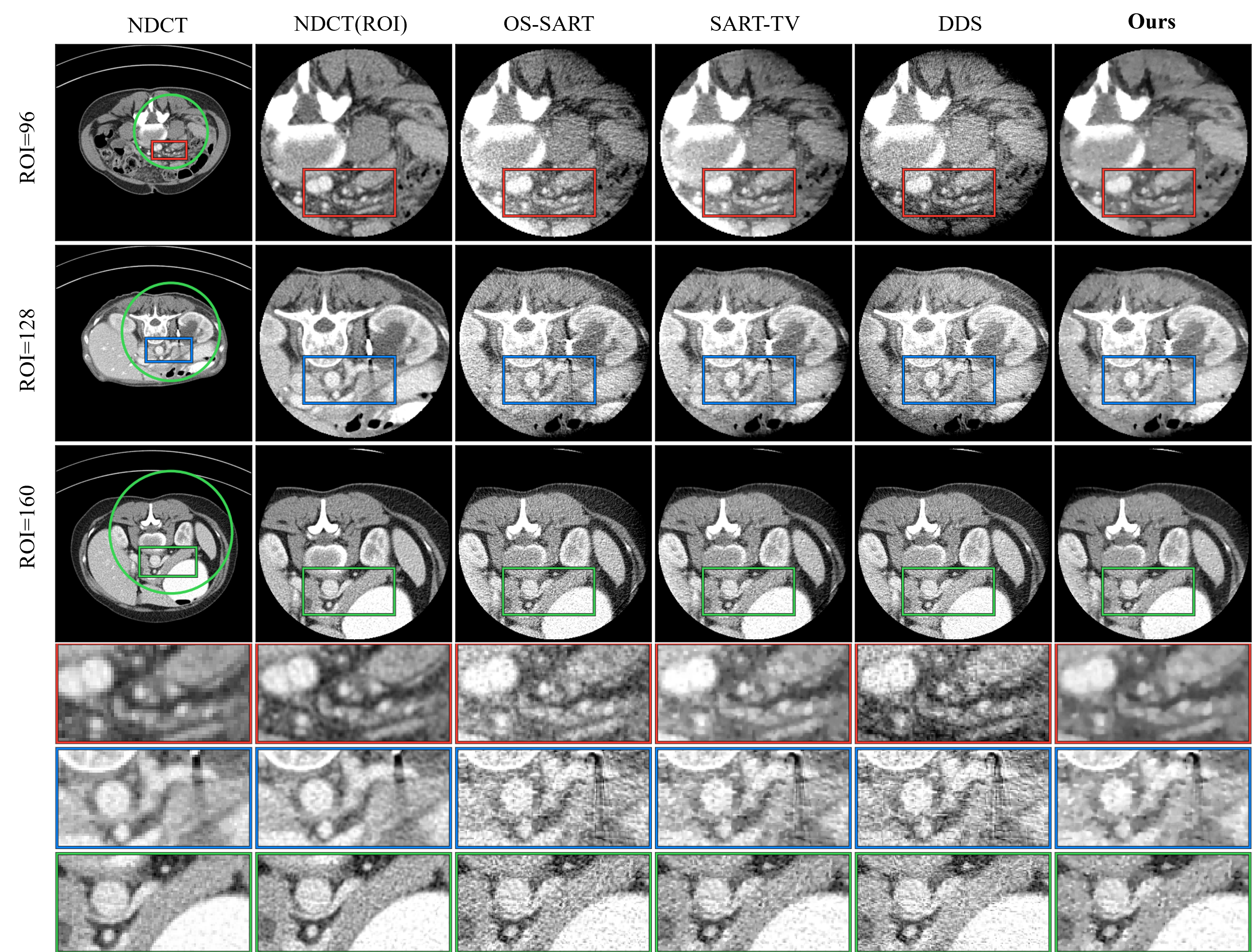}
\caption{Eccentric-ROI reconstructions on real scanner projection low-dose data across ROI radii, soft-tissue window $[-160,240]$~HU. Rows top to bottom: $r{=}96/128/160$~px.}
\label{fig:real_eccentric}
\end{figure*}

\subsubsection{Baselines}
We compare FORCE-Interior with OS-SART~\cite{wang2004ossart}, SART-TV~\cite{sidky2008image,chambolle2004algorithm}, ROI-CT-CNN~\cite{han2019networksolveroisdeep}, and DDS~\cite{chung2024decomposeddiffusionsampleraccelerating} adapted through DM4CT~\cite{shi2026dm4ct}. Hyperparameter details and training-data provenance are reported in Supplementary Sec.~I. ROI-CT-CNN is available only at its trained radius. The recent PEDB method~\cite{wang2025projection} could not be included because no verified implementation was available at the time of this study. A broader comparison with the most recent learning-based methods is left to future work once such implementations become available.

\subsection{Real-Scanner Evaluation}
\label{sec:real}
We first evaluate FORCE-Interior on real scanner projection data. Table~\ref{tab:real_results}, Fig.~\ref{fig:real_centered}, and Fig.~\ref{fig:real_eccentric} report comparisons across ROI radii, positions, and dose levels, with metrics computed against the corresponding vendor reconstruction. FORCE-Interior provides the strongest overall numerical performance, except for a tie with SART-TV in low-dose LPIPS at $r{=}160$. As truncation becomes less severe, its margin generally narrows. These results suggest that the interior-specific design alleviates truncation effects. However, the vendor image is derived from the same examination, is affected by reconstruction and rebinning differences, and is not an independent ground truth.

\subsection{Controlled Synthetic Evaluation}
\label{sec:synthetic}

\begin{table*}[!t]
\centering
\caption{Poisson-noise reconstruction at $r\in\{96,128,160\}$~px (ROI, $[0,2000]$~HU). Best in \textbf{bold}, second best \underline{underlined}. $^{**}$: Holm-adjusted $p<0.001$ vs.\ the best baseline (two-sided paired Wilcoxon, Supplementary Table~S3).}
\label{tab:recon_results}
\resulttablesetup
\begin{tabularx}{\textwidth}{c l *{3}{>{\centering\arraybackslash}X}}
\toprule
\multicolumn{2}{c}{\textit{Experimental setting}}
& \multicolumn{3}{c}{\textit{ROI metrics}} \\
\cmidrule(lr){1-2} \cmidrule(lr){3-5}
ROI radius
& Method
& PSNR(ROI) $\uparrow$
& SSIM(ROI) $\uparrow$
& LPIPS(ROI) $\downarrow$ \\
\midrule

\multirow{4}{*}{$96$}
& OS-SART
& $21.70 \pm 1.48$
& $0.947 \pm 0.009$
& $0.324 \pm 0.055$ \\

& SART-TV
& $21.87 \pm 1.48$
& \underline{$0.953 \pm 0.008$}
& \underline{$0.281 \pm 0.047$} \\

& DDS
& \underline{$28.51 \pm 1.23$}
& $0.948 \pm 0.006$
& $0.352 \pm 0.066$ \\

\rowcolor{ourshl}\cellcolor{white}
& \textbf{Ours}
& \tb{$30.83 \pm 2.01$}$^{**}$
& \tb{$0.973 \pm 0.007$}$^{**}$
& \tb{$0.185 \pm 0.052$}$^{**}$ \\

\cmidrule(lr){1-5}

\multirow{5}{*}{$128$}
& OS-SART
& $29.74 \pm 1.81$
& $0.938 \pm 0.010$
& $0.249 \pm 0.049$ \\

& SART-TV
& $30.00 \pm 1.73$
& \underline{$0.941 \pm 0.007$}
& \underline{$0.218 \pm 0.039$} \\

& ROI-CT-CNN
& $30.11 \pm 1.24$
& $0.936 \pm 0.009$
& $0.264 \pm 0.056$ \\

& DDS
& \underline{$31.16 \pm 2.54$}
& $0.932 \pm 0.021$
& $0.281 \pm 0.091$ \\

\rowcolor{ourshl}\cellcolor{white}
& \textbf{Ours}
& \tb{$33.15 \pm 1.88$}$^{**}$
& \tb{$0.952 \pm 0.016$}$^{**}$
& \tb{$0.191 \pm 0.046$}$^{**}$ \\

\cmidrule(lr){1-5}

\multirow{4}{*}{$160$}
& OS-SART
& $30.60 \pm 1.51$
& $0.879 \pm 0.023$
& $0.266 \pm 0.036$ \\

& SART-TV
& $30.69 \pm 1.58$
& $0.880 \pm 0.022$
& $0.243 \pm 0.035$ \\

& DDS
& \tb{$33.59 \pm 1.24$}
& \tb{$0.926 \pm 0.015$}
& \tb{$0.216 \pm 0.033$} \\

\rowcolor{ourshl}\cellcolor{white}
& \textbf{Ours}
& \underline{$32.57 \pm 2.11$}
& \underline{$0.908 \pm 0.024$}
& \underline{$0.222 \pm 0.047$} \\

\bottomrule
\end{tabularx}
\end{table*}

We then complement the real-scanner validation with a controlled synthetic evaluation that provides a ground truth. Table~\ref{tab:recon_results} reports the quantitative comparison on structure-rich ROIs across ROI radii. Supplementary Fig.~S1 further shows qualitative results at the default radius $r{=}128$. OS-SART recovers the main anatomical structures but leaves substantial noise. SART-TV suppresses noise but tends to oversmooth fine textures. ROI-CT-CNN yields lower ROI metrics and residual artifacts. DDS attains competitive PSNR but lower SSIM and higher LPIPS in these experiments. This pattern may reflect noise amplification by its data-consistency update, although the present comparison does not isolate that mechanism.
Across ROI radii, the qualitative results (Supplementary Fig.~S2) remain consistent with the same trend as in real data evaluations. ROI-CT-CNN is reported only at $r{=}128$ because it was specifically trained with this radius. FORCE-Interior achieves the best PSNR, SSIM, and LPIPS at $r{=}96$ and $r{=}128$. As the ROI grows and truncation becomes less severe, the interior problem becomes better posed. Its advantage is largest at smaller radii, where DDS becomes noticeably blurrier as truncation increases.

\subsection{Measurement Consistency}
\label{sec:hallucination}

\begin{figure}[!t]
\centering
\includegraphics[width=\columnwidth]{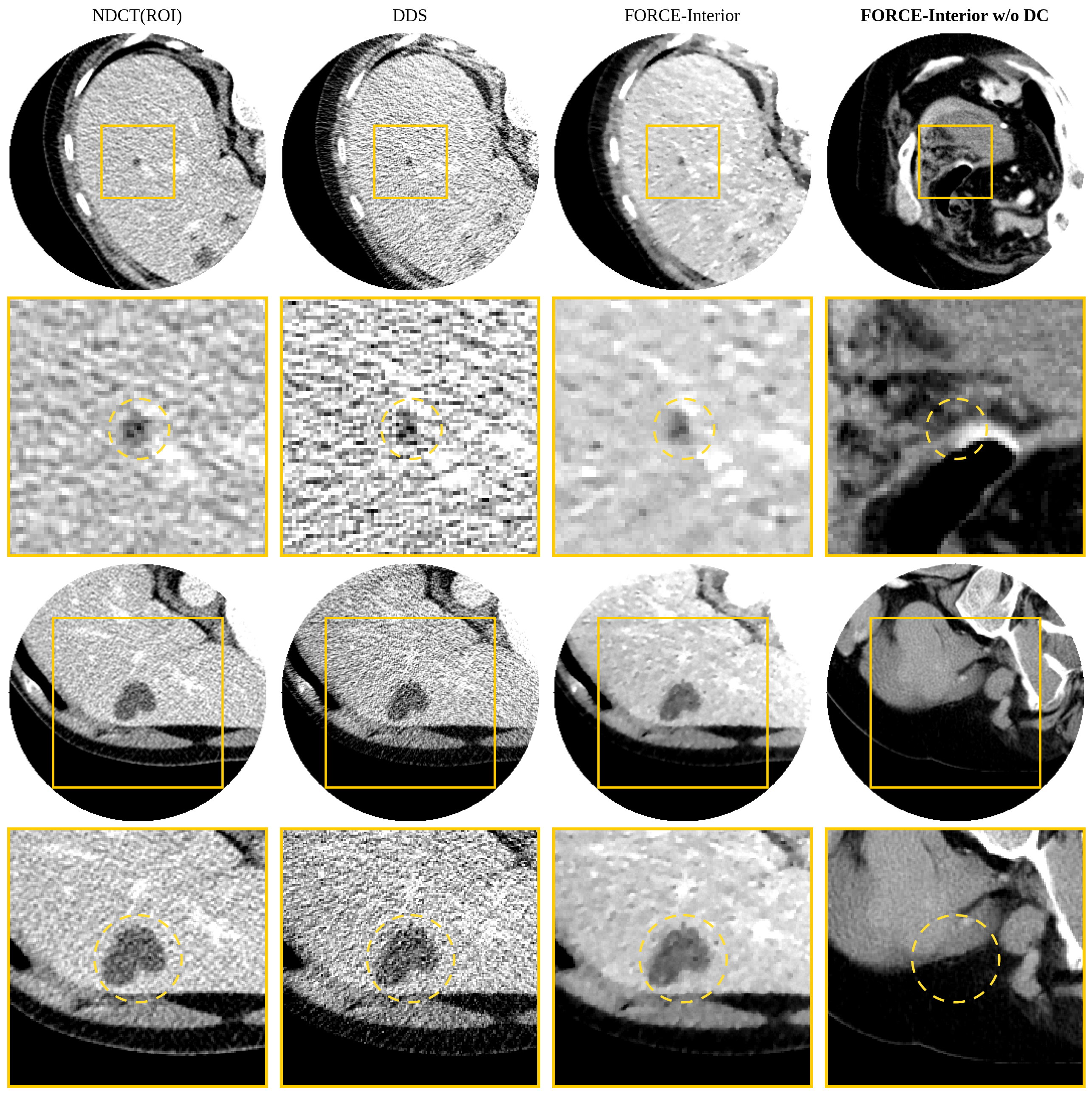}
\caption{Qualitative comparison of lesion preservation.}
\label{fig:real_data_hallu}
\end{figure}

\subsubsection{Lesion Preservation on Real-Scanner Data}
To examine whether generative interior reconstruction preserves clinically relevant structures, we evaluated 20 clinically annotated low-dose slices from the real-scanner dataset using lesion contrast error, CNR, lesion mean-HU error, and local SSIM. We used dataset-provided pathology-location information. Table~\ref{tab:lowdose} shows that FORCE-Interior yields the lowest contrast and mean-HU errors among the evaluated methods and the highest local SSIM. Its CNR exceeds that of the reference, plausibly because background-noise suppression increases visibility. CNR is therefore not treated as a fidelity ranking. These results provide preliminary evidence of lesion preservation relative to the vendor reference, and establishing clinical reliability would require a larger cohort. Removing per-step data consistency markedly increases contrast and HU errors and decreases local SSIM, illustrating that an insufficiently constrained prior can alter lesion attenuation or nearby structures. As shown in Fig.~\ref{fig:real_data_hallu}, the unconstrained reconstruction produces content that does not agree with the reference image.

\begin{table}[!t]
\renewcommand{\arraystretch}{1.2}
\setlength{\tabcolsep}{3pt}
\caption{Real-lesion preservation on low-dose real projections. The first row is the reference and is not ranked. Best in \textbf{bold} among the reconstruction methods, over the three fidelity metrics only. CNR is reported as a lesion-visibility indicator rather than a fidelity metric, because noise suppression can raise it above the reference value.}
\label{tab:lowdose}
\centering
\scriptsize
\begin{tabular}{lcccc}
\toprule
Method & Contr.\,Err.\,$\downarrow$ & CNR & HU\,Err.\,$\downarrow$ & SSIM\,$\uparrow$ \\
\midrule
\textit{NDCT (ROI), reference} & --            & $1.79{\pm}1.27$ & --            & $1.00{\pm}0.00$ \\
\midrule
DDS          & $7.6{\pm}5.1$ & $0.70{\pm}0.50$ & $11.4{\pm}4.5$ & $0.73{\pm}0.10$ \\
FORCE-Interior & $\mathbf{5.6{\pm}6.3}$ & $3.02{\pm}2.50$ & $\mathbf{7.3{\pm}6.0}$ & $\mathbf{0.88{\pm}0.04}$ \\
FORCE-Interior w/o DC & $216.9{\pm}341.0$ & $1.68{\pm}1.67$ & $85.0{\pm}55.3$ & $0.46{\pm}0.18$ \\
\bottomrule
\end{tabular}
\end{table}

\subsubsection{Projection-domain measurement consistency on synthetic data}
\label{sec:hallucination_consistency}
To further assess whether FORCE-Interior reduces the risk of measurement-inconsistent or generative drift, we evaluate projection-domain consistency by forward-projecting the reconstructed image and comparing it with the acquired ROI-truncated sinogram. We use projection-domain residuals as a measurement-consistency test, and further complement this analysis with the small-lesion retention study in Supplementary Sec.~III. Table~\ref{tab:proj_residual} reports the residuals for the structure-rich test subset at ROI radii of 96, 128, and 160 pixels.
FORCE-Interior keeps both residuals small across all ROI radii. Removing per-step data consistency causes a large increase. At $r{=}128$ the measurement residual rises from $0.0080$ to $0.1895$, about $24\times$ higher. Replacing the full-FOV warm start with an ROI-restricted warm start raises it to $0.0125$. Thus, per-step conditioning limits drift from the measured sinogram and the full-FOV warm start further reduces the residual. A small residual establishes measurement consistency, not anatomical uniqueness, because multiple images can agree with truncated data.

\begin{table}[!t]
\centering
\caption{Projection-domain consistency. Lower is better, best in \textbf{bold}.}
\label{tab:proj_residual}
\footnotesize
\setlength{\tabcolsep}{2pt}
\begin{tabular*}{\columnwidth}{@{\extracolsep{\fill}}lccc@{}}
\toprule
Configuration & $r{=}96$~px & $r{=}128$~px & $r{=}160$~px \\
\midrule
\multicolumn{4}{@{}l}{\textit{Measurement residual}}\\
\shortstack[l]{\textbf{FORCE-Interior}}
& \textbf{0.0095\,$\pm$\,0.0011}
& \textbf{0.0080\,$\pm$\,0.0012}
& \textbf{0.0073\,$\pm$\,0.0008} \\
\addlinespace[2pt]
w/o per-step DC
& 0.2897\,$\pm$\,0.0676
& 0.1895\,$\pm$\,0.0580
& 0.1435\,$\pm$\,0.0514 \\
\addlinespace[2pt]
\shortstack[l]{ROI-restricted\\ warm start}
& --
& 0.0125\,$\pm$\,0.0072
& -- \\
\midrule
\multicolumn{4}{@{}l}{\textit{Clean projection residual}}\\
\shortstack[l]{\textbf{FORCE-Interior}}
& \textbf{0.0073\,$\pm$\,0.0012}
& \textbf{0.0055\,$\pm$\,0.0012}
& \textbf{0.0048\,$\pm$\,0.0009} \\
\addlinespace[2pt]
w/o per-step DC
& 0.2895\,$\pm$\,0.0677
& 0.1893\,$\pm$\,0.0580
& 0.1433\,$\pm$\,0.0514 \\
\addlinespace[2pt]
\shortstack[l]{ROI-restricted\\ warm start}
& --
& 0.0109\,$\pm$\,0.0074
& -- \\
\bottomrule
\end{tabular*}
\end{table}

\subsection{Component Analysis}
\subsubsection{Contribution of the generative prior}
We further separate the contribution of the PFGM++ prior from that of the measurement-consistency components. Simply removing the denoiser is ill-posed, because the probability-flow ODE relies on it to estimate the clean image at each step, so the sampling trajectory collapses into noise. We therefore use \emph{Matched-DC-TV}, a deterministic control with the same full-FOV OS-SART warm start and the identical per-step data-consistency and TV schedule as FORCE-Interior, but with no PFGM++ perturbation, denoiser, or probability-flow sampling. It matches the measurement and regularization budget of our method while using no learned prior. As shown in Table~\ref{tab:prior_contrib} ($r{=}128$, normal dose), FORCE-Interior improves the ROI PSNR over Matched-DC-TV by $2.68$~dB (slice-level $95\%$ CI $[2.35, 3.02]$), the SSIM by $0.018$, and the LPIPS by $0.065$, and is better on all $50$ slices. The gain persists under an independent full-FOV reference (PSNR $+2.97$~dB) and holds for all five patients, and Supplementary Sec.~V reports the five-seed averaging used here together with the patient-level analysis. Thus, the generative prior contributes structural and perceptual fidelity beyond what the warm start and repeated data-consistency updates achieve alone.

\begin{table}[!t]
\centering
\caption{Contribution of the generative prior on real scanner data at $r{=}128$ (normal dose, $N{=}50$). $\Delta_{\mathrm{prior}}$ is the mean paired difference favoring FORCE-Interior. Best in \textbf{bold}.}
\label{tab:prior_contrib}
\footnotesize
\setlength{\tabcolsep}{2.5pt}
\begin{tabular*}{\columnwidth}{@{\extracolsep{\fill}}lccc@{}}
\toprule
Method & PSNR $\uparrow$ & SSIM $\uparrow$ & LPIPS $\downarrow$ \\
\midrule
Matched-DC-TV & 29.06 & 0.932 & 0.229 \\
\rowcolor{ourshl}\textbf{FORCE-Interior (Ours)} & \tb{31.75} & \tb{0.950} & \tb{0.164} \\
\midrule
$\Delta_{\mathrm{prior}}$ & $+2.68$ & $+0.018$ & $+0.065$ \\
\bottomrule
\end{tabular*}
\end{table}

\subsubsection{Sampling noise level}
The $t_{\mathrm{start}}$, warm-start, and TV-weight ablations that follow use the $20$-slice radius-sweep subset described in Supplementary Sec.~I rather than the $34$-slice main benchmark.
We also studied the effect of the sampling starting point $t_{\mathrm{start}}$, which controls how much noise is injected into the OS-SART warm start before sampling refinement. A smaller $t_{\mathrm{start}}$ preserves more of the OS-SART initialization, while a larger $t_{\mathrm{start}}$ lets the PFGM++ prior play a stronger role. As shown in Table~\ref{tab:tstart_ablation}, increasing $t_{\mathrm{start}}$ from $0.2$ to $0.6$ steadily improves reconstruction, indicating that the prior plays an important role in refining the warm start. However, at $t_{\mathrm{start}}=0.8$ the quality drops sharply: too much noise is injected and the prior can override the measurement-constrained structure, producing anatomy that is less consistent with the initialization. We therefore use $t_{\mathrm{start}}=0.6$ as the default in all other experiments.

\begin{table}[!t]
\centering
\caption{Effect of $t_{\mathrm{start}}$ at fixed $\lambda{=}0.003$ ($N{=}20$, ROI $[0,2000]$~HU). Best in \textbf{bold}.}
\label{tab:tstart_ablation}
\footnotesize
\setlength{\tabcolsep}{3pt}
\begin{tabular*}{\columnwidth}{@{\extracolsep{\fill}}cccc@{}}
\toprule
$t_{\mathrm{start}}$ & PSNR $\uparrow$ & SSIM $\uparrow$ & LPIPS $\downarrow$ \\
\midrule
0.2 & 31.20 & 0.940 & 0.239 \\
0.4 & 32.05 & 0.942 & 0.228 \\
0.6 & \textbf{32.92} & \textbf{0.953} & \textbf{0.197} \\
0.8 & 30.65 & 0.926 & 0.256 \\
\bottomrule
\end{tabular*}
\end{table}

\subsubsection{Warm-Start}
\label{sec:warm_start_ablation}
\begin{figure}[!t]
\centering
\includegraphics[width=\columnwidth]{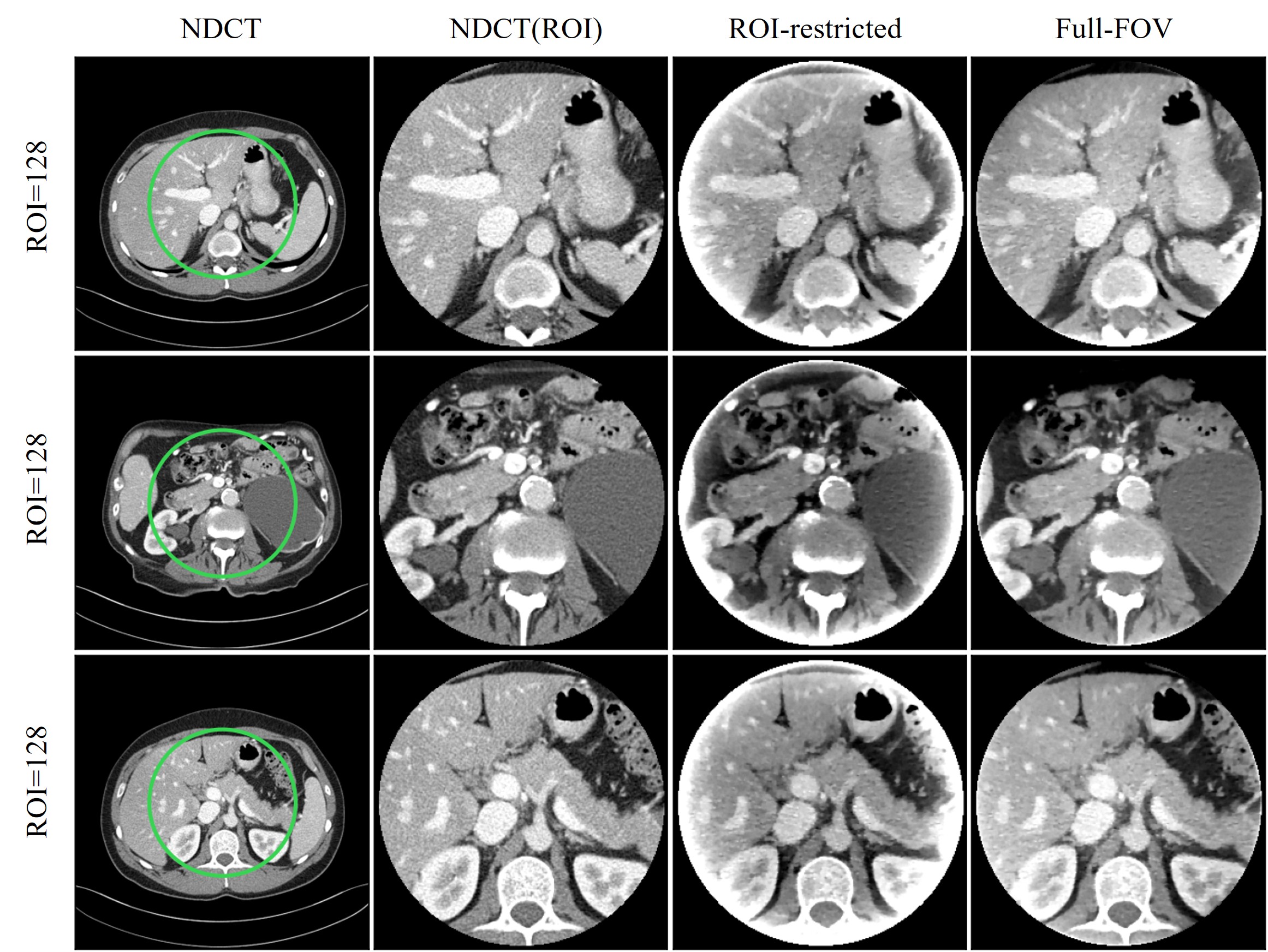}
\caption{Warm-start ablation, window $[-160,240]$~HU.}
\label{fig:warm_start_ablation}
\end{figure}

Table~\ref{tab:init_ablation} compares different warm-start initializations ($t_{\mathrm{start}}=0.6$) to validate our full-FOV warm-start design. As shown in Fig.~\ref{fig:warm_start_ablation} and Table~\ref{tab:init_ablation}, warm starting from full-FOV OS-SART produces better reconstruction results than warm starting from ROI-restricted OS-SART, both quantitatively and qualitatively: the ROI-restricted warm start yields a brighter, washed-out ROI with cupping artifacts and reduced soft-tissue contrast. When out-of-ROI attenuation is not accounted for, the reconstruction develops intensity bias and cupping rather than matching the true attenuation.

\begin{table}[!t]
\centering
\caption{Warm-start ablation at fixed $t_{\mathrm{start}}{=}0.6$ and $\lambda{=}0.003$ ($N{=}20$, ROI $[0,2000]$~HU). Best in \textbf{bold}.}
\label{tab:init_ablation}
\footnotesize
\setlength{\tabcolsep}{2.5pt}
\begin{tabular*}{\columnwidth}{@{\extracolsep{\fill}}lccc@{}}
\toprule
Warm-start initialization & PSNR $\uparrow$ & SSIM $\uparrow$ & LPIPS $\downarrow$ \\
\midrule
ROI-restricted OS-SART & 20.51 & 0.920 & 0.353 \\
Full-FOV OS-SART (\textbf{ours}) & \textbf{32.92} & \textbf{0.953} & \textbf{0.197} \\
\bottomrule
\end{tabular*}
\end{table}

\subsubsection{TV weight}
The total-variation weight $\lambda$ balances noise suppression against detail preservation under noisy acquisition. Table~\ref{tab:tv_ablation} sweeps $\lambda$ from $0.001$ to $0.008$ on the random 20-slice subset. Too small a $\lambda$ under-regularizes and leaves residual noise, which lowers SSIM, whereas too large a $\lambda$ oversmooths and removes fine texture, degrading the perceptual metrics. PSNR stays nearly flat across this range, while SSIM is tied at $0.953$ for $\lambda{=}0.002$ and $0.003$. We adopt $\lambda{=}0.003$ for all experiments, while LPIPS is minimized at $\lambda{=}0.002$.

\begin{table}[!t]
\centering
\caption{Ablation on the TV weight $\lambda$ at fixed $t_{\mathrm{start}}{=}0.6$ ($N{=}20$, ROI $[0,2000]$~HU). Best in \textbf{bold}.}
\label{tab:tv_ablation}
\footnotesize
\setlength{\tabcolsep}{3pt}
\begin{tabular*}{\columnwidth}{@{\extracolsep{\fill}}cccc@{}}
\toprule
$\lambda$ & PSNR $\uparrow$ & SSIM $\uparrow$ & LPIPS $\downarrow$ \\
\midrule
0.001 & 32.40 & 0.949 & 0.195 \\
0.002 & 32.86 & \textbf{0.953} & \textbf{0.177} \\
0.003 & \textbf{32.92} & \textbf{0.953} & 0.197 \\
0.006 & 32.72 & 0.950 & 0.229 \\
0.008 & 32.55 & 0.949 & 0.242 \\
\bottomrule
\end{tabular*}
\end{table}

\section{Discussion and Conclusion}
FORCE-Interior adapts a reusable Poisson-flow CT prior to interior tomography through a full-FOV OS-SART warm start and truncation-mask-aware per-step conditioning. The strongest gains occur under more severe truncation, while DDS becomes competitive at the largest synthetic ROI. The matched deterministic control indicates that the learned prior contributes beyond repeated data consistency and TV alone. Projection residuals establish fidelity to measured rays, but cannot prove anatomical uniqueness in this non-identifiable inverse problem. Lesion-retention experiments provide complementary, still preliminary, evidence.

Several limitations define the scope of these findings. The PFGM++ prior was trained on a modest number of slices from one dataset and may be sensitive to anatomy, reconstruction kernel, scanner, or pathology shifts. Real helical cone-beam data were rebinned to a 2D fan-beam approximation, and the vendor reconstruction is an imperfect reference rather than ground truth. The real cohort contains only five patients. Slice-level tests are exploratory and must be read with the patient-level analysis. Radius-dependent structure-rich selection produces different cohorts, so cross-radius comparisons are descriptive. The uniform-weight OS-SART update approximates rather than models Poisson statistics, and the TV weight and starting noise level require validation beyond the current setting. Finally, a runtime of approximately 42~s per slice is not yet suitable for routine volumetric use.

In conclusion, the results support measurement-consistent, training-free adaptation of a pretrained Poisson-flow prior for interior CT. They do not establish clinical readiness. Broader multi-center patient-level evaluation, matched cohorts across ROI geometries, statistically weighted conditioning, additional contemporary generative baselines, and prospective task-based assessment are needed before clinical deployment.

\bibliographystyle{IEEEtran}
\bibliography{references}
\end{document}